\documentclass[aps,prl,showpacs,twocolumn,groupedaddress,superscriptaddress]{revtex4} % for submission
\usepackage{graphicx}% Include figure files
\usepackage{dcolumn}% Align table columns on decimal point
\usepackage{bm}% bold math
\usepackage{epsfig}
\usepackage{multirow}
\begin{document}

%
%      %%%%%%%%%%%%%%%%%%%%%%%%%%%%%%%%%%%
%%%%%%%%%%%% SOME USEFUL LaTeX2e DEFINITIONS %%%%%%%%%%%%%%
%      %%%%%%%%%%%%%%%%%%%%%%%%%%%%%%%%%%%
%
\newcommand{\comm}[1]{\mbox{\mbox{\textup{#1}}}}
\newcommand{\subs}[1]{\mbox{\scriptstyle \mathit{#1}}}
\newcommand{\subss}[1]{\mbox{\scriptscriptstyle \mathit{#1}}}
\newcommand{\Frac}[2]{\mbox{\frac{\displaystyle{#1}}{\displaystyle{#2}}}}
\newcommand{\LS}[1]{\mbox{_{\scriptstyle \mathit{#1}}}}
\newcommand{\US}[1]{\mbox{^{\scriptstyle \mathit{#1}}}}
\def\gsim{\mathrel{\rlap{\raise.4ex\hbox{$>$}} {\lower.6ex\hbox{$\sim$}}}}
\def\lsim{\mathrel{\rlap{\raise.4ex\hbox{$<$}} {\lower.6ex\hbox{$\sim$}}}}
\renewcommand{\arraystretch}{1.3}
\newcommand{\Edep}{\mbox{E_{\mathit{dep}}}}
\newcommand{\Ebeam}{\mbox{E_{\mathit{beam}}}}
\newcommand{\Exrc}{\mbox{E_{\mathit{rec}}^{\mathit{exp}}}}
\newcommand{\Emrc}{\mbox{E_{\mathit{rec}}^{\mathit{sim}}}}
\newcommand{\Evis}{\mbox{E_{\mathit{vis}}}}
\newcommand{\Edepi}{\mbox{E_{\mathit{dep,i}}}}
\newcommand{\Evisi}{\mbox{E_{\mathit{vis,i}}}}
\newcommand{\Exrci}{\mbox{E_{\mathit{rec,i}}^{\mathit{exp}}}}
\newcommand{\Emrci}{\mbox{E_{\mathit{rec,i}}^{\mathit{sim}}}}
\newcommand{\Etmis}{\mbox{E_{\mathit{t,miss}}}}
%
%%% D0 and the physics analyses related stuff:
%
\newcommand{\lt}{\mbox{$<$}}
\newcommand{\gt}{\mbox{$>$}}
\newcommand{\lte}{\mbox{$\le$}}
\newcommand{\gte}{\mbox{$\ge$}}

\newcommand{\xa}{\mbox{$x_{a}$}}
\newcommand{\xb}{\mbox{$x_{b}$}}
\newcommand{\xp}{\mbox{$x_{p}$}}
\newcommand{\xpb}{\mbox{$x_{\bar{p}}$}}

\newcommand{\alphas}{\mbox{$\alpha_{s}$}}
\newcommand{\partt}{\mbox{$\partial^{t}$}}
\newcommand{\partd}{\mbox{$\partial_{t}$}}
\newcommand{\parmut}{\mbox{$\partial^{\mu}$}}
\newcommand{\parmud}{\mbox{$\partial_{\mu}$}}
\newcommand{\parnut}{\mbox{$\partial^{\nu}$}}
\newcommand{\parnud}{\mbox{$\partial_{\nu}$}}
\newcommand{\Amut}{\mbox{$A^{\mu}$}}
\newcommand{\Amud}{\mbox{$A_{\mu}$}}
\newcommand{\Anut}{\mbox{$A^{\nu}$}}
\newcommand{\Anud}{\mbox{$A_{\nu}$}}
\newcommand{\Fmunut}{\mbox{$F^{\mu\nu}$}}
\newcommand{\Fmunud}{\mbox{$F_{\mu\nu}$}}
\newcommand{\Gmut}{\mbox{$\gamma^{\mu}$}}
\newcommand{\Gmud}{\mbox{$\gamma_{\mu}$}}
\newcommand{\Gnut}{\mbox{$\gamma^{\nu}$}}
\newcommand{\Gnud}{\mbox{$\gamma_{\nu}$}}
\newcommand{\pvecsq}{\mbox{$\vec{p}^{\:2}$}}
\newcommand{\bigsum}{\mbox{$\displaystyle{\sum}$}}
%
% Triggers:
\newcommand{\Lzero}{Level \O}
\newcommand{\Lone}{Level $1$}
\newcommand{\Ltwo}{Level $2$}
\newcommand{\Lhalf}{Level $1.5$}
%
% D0 General
\newcommand{\bzero}{\mbox{\comm{B\O}}}
\newcommand{\dzero}{\mbox{D\O}}
\newcommand{\dzerosm}{\mbox{\comm{$\scriptsize{D\O}$}}}
\newcommand{\runb}{Run~$1$B}
\newcommand{\runa}{Run~$1$A}
\newcommand{\runone}{Run~$1$}
\newcommand{\runtwo}{Run~$2$}
\newcommand{\dzpjet}{\textsc{D{\O}Pjet}}
\newcommand{\y}{\mbox{$y$}}
\newcommand{\z}{\mbox{$z$}}
\newcommand{\px}{\mbox{$p_{x}$}}
\newcommand{\py}{\mbox{$p_{y}$}}
\newcommand{\pz}{\mbox{$p_{z}$}}
\newcommand{\ex}{\mbox{$E_{x}$}}
\newcommand{\ey}{\mbox{$E_{y}$}}
\newcommand{\ez}{\mbox{$E_{z}$}}
\newcommand{\et}{\mbox{$E_{T}$}}
\newcommand{\etprime}{\mbox{$E_{T}^{\prime}$}}
\newcommand{\etone}{\mbox{$E_{T}^{\mathrm{1}}$}}
\newcommand{\ettwo}{\mbox{$E_{T}^{\mathrm{2}}$}}
\newcommand{\etlj}{\mbox{$E_{T}^{\subs{lj}}$}}
\newcommand{\etmax}{\mbox{$E_{T}^{max}$}}
\newcommand{\etcand}{\mbox{$E_{T}^{\subs{cand}}$}}
\newcommand{\etup}{\mbox{$E_{T}^{\subs{up}}$}}
\newcommand{\etdown}{\mbox{$E_{T}^{\subs{down}}$}}
\newcommand{\jet}{\mbox{$E_{T}^{\subs{jet}}$}}
\newcommand{\cet}{\mbox{$E_{T}^{\subs{cell}}$}}
\newcommand{\jetvec}{\mbox{$\vec{E}_{T}^{\subs{jet}}$}}
\newcommand{\cetvec}{\mbox{$\vec{E}_{T}^{\subs{cell}}$}}
\newcommand{\jevec}{\mbox{$\vec{E}^{\subs{jet}}$}}
\newcommand{\cevec}{\mbox{$\vec{E}^{\subs{cell}}$}}
\newcommand{\etfr}{\mbox{$f_{E_{T}}$}}
\newcommand{\aveet}{\mbox{$\langle\et\rangle$}}
\newcommand{\nj}{\mbox{$n_{j}$}}
\newcommand{\ptrel}{\mbox{$p_{T}^{rel}$}}
\newcommand{\etad}{\mbox{$\eta_{d}$}}           %% detector eta
\newcommand{\peta}{\mbox{$\eta$}}               %% physics eta
\newcommand{\aeta}{\mbox{$|\eta|$}}             %% absolute eta
\newcommand{\ifb}{fb$^{-1}$}
\newcommand{\ipb}{pb$^{-1}$}
\newcommand{\inb}{nb$^{-1}$}
\newcommand{\met}{\mbox{${\hbox{$E$\kern-0.63em\lower-.18ex\hbox{/}}}_{T}$}}
\newcommand{\metvec}{\mbox{${\hbox{$\vec{E}$\kern-0.63em\lower-.18ex\hbox{/}}}_{T}\,$}}
\newcommand{\metx}{\mbox{${\hbox{$E$\kern-0.63em\lower-.18ex\hbox{/}}}_{x}\,$}}
\newcommand{\mety}{\mbox{${\hbox{$E$\kern-0.63em\lower-.18ex\hbox{/}}}_{y}\,$}}
\newcommand{\het}{\mbox{$\vec{\mathcal{H}}_{T}$}}
\newcommand{\hetsc}{\mbox{$\mathcal{H}_{T}$}}
\newcommand{\zvrt}{\mbox{$Z$}}
\newcommand{\zcut}{\mbox{$|\zvrt| < 50$}}
\newcommand{\mtwo}{\mbox{$\mathcal{M}_{2}$}}
\newcommand{\mthree}{\mbox{$\mathcal{M}_{3}$}}
\newcommand{\mfour}{\mbox{$\mathcal{M}_{4}$}}
\newcommand{\msix}{\mbox{$\mathcal{M}_{6}$}}
\newcommand{\mn}{\mbox{$\mathcal{M}_{n}$}}
\newcommand{\R}{\mbox{$R_{\subss{MTE}}$}}
\newcommand{\invR}{\mbox{$1/R_{\subss{MTE}}$}}
\newcommand{\eemf}{\mbox{$\varepsilon_{\subss{EMF}}$}}
\newcommand{\echf}{\mbox{$\varepsilon_{\subss{CHF}}$}}
\newcommand{\ehcf}{\mbox{$\varepsilon_{\subss{HCF}}$}}
\newcommand{\eglob}{$\mbox{\varepsilon_{\subs{glob}}$}}
\newcommand{\emte}{\mbox{$\varepsilon_{\subss{MTE}}$}}
\newcommand{\ezvrt}{\mbox{$\varepsilon_{\subss{Z}}$}}
\newcommand{\etot}{\mbox{$\varepsilon_{\subs{tot}}$}}
\newcommand{\Bprime}{\mbox{$\comm{B}^{\prime}$}}
\newcommand{\Nsurv}{\mbox{$N_{\subs{surv}}$}}
\newcommand{\Nfail}{\mbox{$N_{\subs{fail}}$}}
\newcommand{\Ntot}{\mbox{$N_{\subs{tot}}$}}
\newcommand{\p}[1]{\mbox{$p_{#1}$}}
\newcommand{\ep}[1]{\mbox{$\Delta p_{#1}$}}
\newcommand{\delr}{\mbox{$\Delta R$}}
\newcommand{\deleta}{\mbox{$\Delta\eta$}}
\newcommand{\cafone}{{\sc Cafix 5.1}}
\newcommand{\caftwo}{{\sc Cafix 5.2}}
\newcommand{\delphi}{\mbox{$\Delta\varphi$}}
\newcommand{\rphi}{\mbox{$r-\varphi$}}
\newcommand{\etaphi}{\mbox{$\eta-\varphi$}}
\newcommand{\etatphi}{\mbox{$\eta\times\varphi$}}
\newcommand{\Rjet}{\mbox{$R_{jet}$}}
\newcommand{\jphi}{\mbox{$\varphi_{\subs{jet}}$}}
\newcommand{\gphi}{\mbox{$\varphi_{\subs{\gamma}}$}}
\newcommand{\ceta}{\mbox{$\eta^{\subs{cell}}$}}
\newcommand{\cphi}{\mbox{$\phi^{\subs{cell}}$}}
\newcommand{\inlum}{\mbox{$\mathcal{L}$}}
\newcommand{\gm}{\mbox{$\gamma$}}
%\newcommand{\Rjj}{\mbox{\mathcal{R}_{\subs{jj}}}}
% for thesis use this instead:
\newcommand{\Rjj}{\mbox{$\mathbf{R}_{\subs{jj}}$}}
\newcommand{\Rgj}{\mbox{$\mathbf{R}_{\subs{\gamma j}}$}}
\newcommand{\Rmathcal}{\mbox{$\mathcal{R}$}}
\newcommand{\etv}{\mbox{$\vec{E}_{T}$}}
\newcommand{\nvec}{\mbox{$\hat{\vec{n}}$}}
\newcommand{\eprime}{\mbox{$E^{\prime}$}}
\newcommand{\aveprime}{\mbox{$\bar{E}^{\prime}$}}
\newcommand{\geta}{\mbox{$\eta_{\gm}$}}
\newcommand{\jeta}{\mbox{$\eta_{\subs{jet}}$}}
\newcommand{\cjeta}{\mbox{$\eta_{\subs{jet}}^{\subss{CEN}}$}}
\newcommand{\fjeta}{\mbox{$\eta_{\subs{jet}}^{\subss{FOR}}$}}
\newcommand{\cjphi}{\mbox{$\varphi_{\subs{jet}}^{\subss{CEN}}$}}
\newcommand{\fjphi}{\mbox{$\varphi_{\subs{jet}}^{\subss{FOR}}$}}
\newcommand{\etcut}{\mbox{$E_{T}^{\subs{cut}}$}}
%\newcommand{\etg}{\mbox{E_{T\gm}}}
% for thesis, use this instead.
\newcommand{\etg}{\mbox{$E_{T}^{\gm}$}}
\newcommand{\cenet}{\mbox{$E_{T}^{\subss{CEN}}$}}
\newcommand{\foret}{\mbox{$E_{T}^{\subss{FOR}}$}}
\newcommand{\cenen}{\mbox{$E^{\subss{CEN}}$}}
\newcommand{\foren}{\mbox{$E^{\subss{FOR}}$}}
\newcommand{\ejtptc}{\mbox{$E^{\subs{ptcl}}_{\subs{jet}}$}}
\newcommand{\ejtmes}{\mbox{$E^{\subs{meas}}_{\subs{jet}}$}}
\newcommand{\AIDA}{{\sc AIDA}}
\newcommand{\RECO}{{\sc Reco}}
\newcommand{\PYTHIA}{{\sc Pythia}}
\newcommand{\HERWIG}{{\sc Herwig}}
\newcommand{\pythia}    {\mbox{\sc{pythia}}}
\newcommand{\alpgen}    {\mbox{\sc{alpgen}}}
\newcommand{\mcatnlo}    {\mbox{\sc{mc@nlo}}}
\newcommand{\herwig}    {\mbox{\sc{herwig}}}
\newcommand{\madgraph}    {\mbox{\sc{MadGraph}}}
\newcommand{\GEANT}    {\mbox{\sc{Geant}}}
\newcommand{\JETRAD}{{\sc Jetrad}}
\newcommand{\CTone}{\mbox{$|\eta|<0.4}$}
\newcommand{\CTtwo}{\mbox{$0.4\leq|\eta|<0.8$}}
\newcommand{\ICone}{\mbox{$0.8\leq|\eta|<1.2$}}
\newcommand{\ICtwo}{\mbox{$1.2\leq|\eta|<1.6$}}
\newcommand{\FWone}{\mbox{$1.6\leq|\eta|<2.0$}}
\newcommand{\FWtwo}{\mbox{$2.0\leq|\eta|<2.5$}}
\newcommand{\FWthr}{\mbox{$2.5\leq|\eta|<3.0$}}
\newcommand{\LCTone}{\mbox{$|\eta|<0.5$}}
\newcommand{\LCTtwo}{\mbox{$0.5\leq|\eta|<1.0$}}
\newcommand{\LICone}{\mbox{$1.0\leq|\eta|<1.5$}}
\newcommand{\LICtwo}{\mbox{$1.5\leq|\eta|<2.0$}}
\newcommand{\LFWone}{\mbox{$2.0\leq|\eta|<3.0$}}
\newcommand{\CSone}{\mbox{$|\eta|<0.5$}}
\newcommand{\CStwo}{\mbox{$0.5\leq|\eta|<1.0$}}
\newcommand{\CSthr}{\mbox{$1.0\leq|\eta|<1.5$}}
\newcommand{\CSfou}{\mbox{$1.5\leq|\eta|<2.0$}}
\newcommand{\CSfiv}{\mbox{$2.0\leq|\eta|<3.0$}}
\newcommand{\sigA}{\mbox{$\sigma_{\subss{\!A}}$}}
\newcommand{\sigASS}{\mbox{$\sigma_{\subss{A}}^{\subss{SS}}$}}
\newcommand{\sigAOS}{\mbox{$\sigma_{\subss{A}}^{\subss{OS}}$}}
\newcommand{\sigZ}{\mbox{$\sigma_{\subss{Z}}$}}
\newcommand{\sige}{\mbox{$\sigma_{\subss{E}}$}}
\newcommand{\siget}{\mbox{$\sigma_{\subss{\et}}$}}
\newcommand{\sigetone}{\mbox{$\sigma_{\subs{\etone}}$}}
\newcommand{\sigettwo}{\mbox{$\sigma_{\subs{\ettwo}}$}}
\newcommand{\rcal}{\mbox{$R_{\subs{cal}}$}}
\newcommand{\zcal}{\mbox{$Z_{\subs{cal}}$}}
\newcommand{\Runf}{\mbox{$R_{\subs{unf}}$}}
\newcommand{\Rsep}{\mbox{$\mathcal{R}_{sep}$}}
\newcommand{\etal}{{\it et al.}}
\newcommand{\ppbar}{\mbox{$p\overline{p}$}}
\newcommand{\pp}{\mbox{$pp$}}
\newcommand{\qqbar}{\mbox{$q\overline{q}$}}
\newcommand{\ccbar}{\mbox{$c\overline{c}$}}
\newcommand{\bbbar}{\mbox{$b\overline{b}$}}
\newcommand{\ttbar}{\mbox{$t\overline{t}$}}

\newcommand{\bbj}{\mbox{$b\overline{b}j$}}
\newcommand{\bbjj}{\mbox{$b\overline{b}jj$}}
\newcommand{\ccjj}{\mbox{$c\overline{c}jj$}}

\newcommand{\bb}{\mbox{$b\overline{b}j(j)$}}
\newcommand{\cc}{\mbox{$c\overline{c}j(j)$}}

\newcommand{\hboson}{\mbox{$\mathit{h}$}}
\newcommand{\Hboson}{\mbox{$\mathit{H}$}}
\newcommand{\Aboson}{\mbox{$\mathit{A}$}}
\newcommand{\zboson}{\mbox{$\mathit{Z}$}}
\newcommand{\zb}{\mbox{$\mathit{Zb}$}}
\newcommand{\bh}{\mbox{$\mathit{bh}$}}
\newcommand{\btag}{\mbox{$\mathit{b}$}}

\newcommand{\hsm}{\mbox{$h_{SM}$}}
\newcommand{\hmssm}{\mbox{$h_{MSSM}$}}

\newcommand{\prot}{\mbox{$p$}}
\newcommand{\pbar}{\mbox{$\overline{p}$}}
\newcommand{\pt}{\mbox{$p_{T}$}}
\newcommand{\xnot}{\mbox{$X_{0}$}}
\newcommand{\Znot}{\mbox{$Z^{0}$}}
\newcommand{\Wpm}{\mbox{$W^{\pm}$}}
\newcommand{\Wplus}{\mbox{$W^{+}$}}
\newcommand{\Wminus}{\mbox{$W^{-}$}}
\newcommand{\lamb}{\mbox{$\lambda$}}
\newcommand{\nhatbf}{\mbox{$\hat{\mathbf{n}}$}}
\newcommand{\pbf}{\mbox{$\mathbf{p}$}}
\newcommand{\xbf}{\mbox{$\mathbf{x}$}}
\newcommand{\jbf}{\mbox{$\mathbf{j}$}}
\newcommand{\Ebf}{\mbox{$\mathbf{E}$}}
\newcommand{\Bbf}{\mbox{$\mathbf{B}$}}
\newcommand{\Abf}{\mbox{$\mathbf{A}$}}
\newcommand{\Rbf}{\mbox{$\mathbf{R}$}}
\newcommand{\nablabf}{\mbox{$\mathbf{\nabla}$}}
\newcommand{\rarrow}{\mbox{$\rightarrow$}}
\newcommand{\slashp}{\mbox{$\not \! p \,$}}
\newcommand{\slashk}{\mbox{$\not \! k$}}
\newcommand{\slasha}{\mbox{$\not \! a$}}
\newcommand{\slashA}{\mbox{$\! \not \! \! A$}}
\newcommand{\slashpar}{\mbox{$\! \not \! \partial$}}
\newcommand{\intdouble}{\mbox{$\int\!\!\int$}}
\newcommand{\MRSTGU}{MRSTg$\uparrow$}
\newcommand{\MRSTGD}{MRSTg$\downarrow$}
%
% JES:
\newcommand{\Due}{\mbox{$D_{\mathrm{ue}}$}}
\newcommand{\Dth}{\mbox{$D_{\Theta}$}}
\newcommand{\Dof}{\mbox{$D_{\mathrm{O}}$}}
\newcommand{\zbl}{\texttt{ZERO BIAS}}
\newcommand{\mbl}{\texttt{MIN BIAS}}
\newcommand{\mbll}{\texttt{MINIMUM BIAS}}
\newcommand{\nue}{\mbox{$\nu_{e}$}}
\newcommand{\num}{\mbox{$\nu_{\mu}$}}
\newcommand{\nut}{\mbox{$\nu_{\tau}$}}
\newcommand{\mycs}{\mbox{$d^{\,2}\sigma/(d\et d\eta)$}}
\newcommand{\mycsav}{\mbox{$\langle \mycs \rangle$}}
\newcommand{\tdcs}{\mbox{$d^{\,3}\sigma/d\et d\eta_{1} d\eta_{2}$}}
\newcommand{\tdcsav}{\mbox{$\langle d^{\,3}\sigma/d\et d\eta_{1} d\eta_{2} \rangle$}}
\newcommand{\tanb}{$\tan\beta$}
\newcommand{\cotb}{$\cot\beta$}
%%

% Inclusive Definitions
\newcommand{\rstev}{\mbox{$\rs = \T{1.8}$}}
\newcommand{\rssps}{\mbox{$\rs = \T{0.63}$}}
\newcommand{\XX}{\mbox{$\, \times \,$}}
\newcommand{\AP}{\mbox{${\rm \bar{p}}$}}
\newcommand{\SU}{\mbox{$<\! |S|^2 \!>$}}
\newcommand{\ET}{\mbox{$E_{T}$}}
\newcommand{\HT}{\mbox{$S_{{\rm {\sl T}}}$} }
\newcommand{\PT}{\mbox{$p_{t}$}}
\newcommand{\DP}{\mbox{$\Delta\phi$}}
\newcommand{\DR}{\mbox{$\Delta R$}}
\newcommand{\DE}{\mbox{$\Delta\eta$}}
\newcommand{\DEP}{\mbox{$\Delta\eta_{c}$}}
\newcommand{\PH}{\mbox{$\phi$}}
\newcommand{\EA}{\mbox{$\eta$} }
\newcommand{\EAJ}{\mbox{\EA(jet)}}
\newcommand{\AEA}{\mbox{$|\eta|$}}
\newcommand{\Ge}[1]{\mbox{#1 GeV}}
\newcommand{\T}[1]{\mbox{#1 TeV}}
\newcommand{\x}{\cdot}
\newcommand{\ra}{\rightarrow}
\def\D0{D\O}
\def\ETmiss{{\rm {\mbox{$E\kern-0.57em\raise0.19ex\hbox{/}_{T}$}}}}
% units
\newcommand{\mb}{\mbox{mb}}
\newcommand{\nb}{\mbox{nb}}
\newcommand{\rs}{\mbox{$\sqrt{\rm {\sl s}}$}}
\newcommand{\fdel}{\mbox{$f(\DEP)$}}
\newcommand{\fdele}{\mbox{$f(\DEP)^{exp}$}}
\newcommand{\fgap}{\mbox{$f(\DEP\! \geq \!3)$}}
\newcommand{\fgape}{\mbox{$f(\DEP\! \geq \!3)^{exp}$}}
\newcommand{\fpyt}{\mbox{$f(\DEP\!>\!2)$}}
\newcommand{\delth}{\mbox{$\DEP\! \geq \!3$}}
\newcommand{\uplim}{\mbox{$1.1\!\times\!10^{-2}$}}
\def\simge
{\mathrel{\rlap{\raise 0.53ex \hbox{$>$}}{\lower 0.53ex \hbox{$\sim$}}}}
\def\simle
{\mathrel{\rlap{\raise 0.53ex \hbox{$<$}}{\lower 0.53ex \hbox{$\sim$}}}}
% End Inclusive Definitions
%%
\newcommand{\pbarp}{\mbox{$p\bar{p}$}}
\def\ETmiss{\mbox{${\hbox{$E$\kern-0.5em\lower-.1ex\hbox{/}\kern+0.15em}}_{\rm T}$}}
\def\Et{\mbox{$E_{T}$}}
\newcommand{\modeta}{\mid \!\! \eta \!\! \mid}
\def\gevcc{GeV/c$^2$}                   %GeV/c^2
\def\gevc{GeV/c}                        %GeV/c
\def\gev{GeV}                           %GeV
\newcommand{\als}{\mbox{${\alpha_{{\rm s}}}$}}
\def\1960{$\sqrt{s}=1960$ GeV}
\def\etI{E_{T_1}}
\def\etII{E_{T_2}}
\def\itaI{\eta_1}
\def\itaII{\eta_2}
\def\deta{\Delta\eta}
\def\etab{\bar{\eta}}
\def\xq{($x_1$,$x_2$,$Q^2$)}
\def\xx{($x_1$,$x_2$)}
\def\rap{pseudorapidity}
\def\as{\alpha_s}
\def\ap{\alpha_{\rm BFKL}}
\def\apb{\alpha_{{\rm BFKL}_{bin}}}
\def\cm{c.m.}

\def\MET{{\mbox{$E\kern-0.57em\raise0.19ex\hbox{/}_{T}$}}}

\hspace{5.2in} \mbox{FERMILAB-PUB-11-002-E} %the fermi-preprint number

\title{Measurement of color flow in $\mathbf{t\bar{t}}$ events from $\mathbf{p\bar{p}}$ collisions at $\mathbf{\sqrt{s}=1.96}$ TeV}

% remove these 3 lines before journal submittal.
%\centerline{author list dated 2 December 2010}
% end removal before journal submittal
%
\affiliation{Universidad de Buenos Aires, Buenos Aires, Argentina}
\affiliation{LAFEX, Centro Brasileiro de Pesquisas F{\'\i}sicas, Rio de Janeiro, Brazil}
\affiliation{Universidade do Estado do Rio de Janeiro, Rio de Janeiro, Brazil}
\affiliation{Universidade Federal do ABC, Santo Andr\'e, Brazil}
\affiliation{Instituto de F\'{\i}sica Te\'orica, Universidade Estadual Paulista, S\~ao Paulo, Brazil}
\affiliation{Simon Fraser University, Vancouver, British Columbia, and York University, Toronto, Ontario, Canada}
\affiliation{University of Science and Technology of China, Hefei, People's Republic of China}
\affiliation{Universidad de los Andes, Bogot\'{a}, Colombia}
\affiliation{Charles University, Faculty of Mathematics and Physics, Center for Particle Physics, Prague, Czech Republic}
\affiliation{Czech Technical University in Prague, Prague, Czech Republic}
\affiliation{Center for Particle Physics, Institute of Physics, Academy of Sciences of the Czech Republic, Prague, Czech Republic}
\affiliation{Universidad San Francisco de Quito, Quito, Ecuador}
\affiliation{LPC, Universit\'e Blaise Pascal, CNRS/IN2P3, Clermont, France}
\affiliation{LPSC, Universit\'e Joseph Fourier Grenoble 1, CNRS/IN2P3, Institut National Polytechnique de Grenoble, Grenoble, France}
\affiliation{CPPM, Aix-Marseille Universit\'e, CNRS/IN2P3, Marseille, France}
\affiliation{LAL, Universit\'e Paris-Sud, CNRS/IN2P3, Orsay, France}
\affiliation{LPNHE, Universit\'es Paris VI and VII, CNRS/IN2P3, Paris, France}
\affiliation{CEA, Irfu, SPP, Saclay, France}
\affiliation{IPHC, Universit\'e de Strasbourg, CNRS/IN2P3, Strasbourg, France}
\affiliation{IPNL, Universit\'e Lyon 1, CNRS/IN2P3, Villeurbanne, France and Universit\'e de Lyon, Lyon, France}
\affiliation{III. Physikalisches Institut A, RWTH Aachen University, Aachen, Germany}
\affiliation{Physikalisches Institut, Universit{\"a}t Freiburg, Freiburg, Germany}
\affiliation{II. Physikalisches Institut, Georg-August-Universit{\"a}t G\"ottingen, G\"ottingen, Germany}
\affiliation{Institut f{\"u}r Physik, Universit{\"a}t Mainz, Mainz, Germany}
\affiliation{Ludwig-Maximilians-Universit{\"a}t M{\"u}nchen, M{\"u}nchen, Germany}
\affiliation{Fachbereich Physik, Bergische Universit{\"a}t Wuppertal, Wuppertal, Germany}
\affiliation{Panjab University, Chandigarh, India}
\affiliation{Delhi University, Delhi, India}
\affiliation{Tata Institute of Fundamental Research, Mumbai, India}
\affiliation{University College Dublin, Dublin, Ireland}
\affiliation{Korea Detector Laboratory, Korea University, Seoul, Korea}
\affiliation{CINVESTAV, Mexico City, Mexico}
\affiliation{FOM-Institute NIKHEF and University of Amsterdam/NIKHEF, Amsterdam, The Netherlands}
\affiliation{Radboud University Nijmegen/NIKHEF, Nijmegen, The Netherlands}
\affiliation{Joint Institute for Nuclear Research, Dubna, Russia}
\affiliation{Institute for Theoretical and Experimental Physics, Moscow, Russia}
\affiliation{Moscow State University, Moscow, Russia}
\affiliation{Institute for High Energy Physics, Protvino, Russia}
\affiliation{Petersburg Nuclear Physics Institute, St. Petersburg, Russia}
\affiliation{Stockholm University, Stockholm and Uppsala University, Uppsala, Sweden }
\affiliation{Lancaster University, Lancaster LA1 4YB, United Kingdom}
\affiliation{Imperial College London, London SW7 2AZ, United Kingdom}
\affiliation{The University of Manchester, Manchester M13 9PL, United Kingdom}
\affiliation{University of Arizona, Tucson, Arizona 85721, USA}
\affiliation{University of California Riverside, Riverside, California 92521, USA}
\affiliation{Florida State University, Tallahassee, Florida 32306, USA}
\affiliation{Fermi National Accelerator Laboratory, Batavia, Illinois 60510, USA}
\affiliation{University of Illinois at Chicago, Chicago, Illinois 60607, USA}
\affiliation{Northern Illinois University, DeKalb, Illinois 60115, USA}
\affiliation{Northwestern University, Evanston, Illinois 60208, USA}
\affiliation{Indiana University, Bloomington, Indiana 47405, USA}
\affiliation{Purdue University Calumet, Hammond, Indiana 46323, USA}
\affiliation{University of Notre Dame, Notre Dame, Indiana 46556, USA}
\affiliation{Iowa State University, Ames, Iowa 50011, USA}
\affiliation{University of Kansas, Lawrence, Kansas 66045, USA}
\affiliation{Kansas State University, Manhattan, Kansas 66506, USA}
\affiliation{Louisiana Tech University, Ruston, Louisiana 71272, USA}
\affiliation{Boston University, Boston, Massachusetts 02215, USA}
\affiliation{Northeastern University, Boston, Massachusetts 02115, USA}
\affiliation{University of Michigan, Ann Arbor, Michigan 48109, USA}
\affiliation{Michigan State University, East Lansing, Michigan 48824, USA}
\affiliation{University of Mississippi, University, Mississippi 38677, USA}
\affiliation{University of Nebraska, Lincoln, Nebraska 68588, USA}
\affiliation{Rutgers University, Piscataway, New Jersey 08855, USA}
\affiliation{Princeton University, Princeton, New Jersey 08544, USA}
\affiliation{State University of New York, Buffalo, New York 14260, USA}
\affiliation{Columbia University, New York, New York 10027, USA}
\affiliation{University of Rochester, Rochester, New York 14627, USA}
\affiliation{State University of New York, Stony Brook, New York 11794, USA}
\affiliation{Brookhaven National Laboratory, Upton, New York 11973, USA}
\affiliation{Langston University, Langston, Oklahoma 73050, USA}
\affiliation{University of Oklahoma, Norman, Oklahoma 73019, USA}
\affiliation{Oklahoma State University, Stillwater, Oklahoma 74078, USA}
\affiliation{Brown University, Providence, Rhode Island 02912, USA}
\affiliation{University of Texas, Arlington, Texas 76019, USA}
\affiliation{Southern Methodist University, Dallas, Texas 75275, USA}
\affiliation{Rice University, Houston, Texas 77005, USA}
\affiliation{University of Virginia, Charlottesville, Virginia 22901, USA}
\affiliation{University of Washington, Seattle, Washington 98195, USA}
\author{V.M.~Abazov} \affiliation{Joint Institute for Nuclear Research, Dubna, Russia}
\author{B.~Abbott} \affiliation{University of Oklahoma, Norman, Oklahoma 73019, USA}
\author{B.S.~Acharya} \affiliation{Tata Institute of Fundamental Research, Mumbai, India}
\author{M.~Adams} \affiliation{University of Illinois at Chicago, Chicago, Illinois 60607, USA}
\author{T.~Adams} \affiliation{Florida State University, Tallahassee, Florida 32306, USA}
\author{G.D.~Alexeev} \affiliation{Joint Institute for Nuclear Research, Dubna, Russia}
\author{G.~Alkhazov} \affiliation{Petersburg Nuclear Physics Institute, St. Petersburg, Russia}
\author{A.~Alton$^{a}$} \affiliation{University of Michigan, Ann Arbor, Michigan 48109, USA}
\author{G.~Alverson} \affiliation{Northeastern University, Boston, Massachusetts 02115, USA}
\author{G.A.~Alves} \affiliation{LAFEX, Centro Brasileiro de Pesquisas F{\'\i}sicas, Rio de Janeiro, Brazil}
% ...
%
% visitor_addresses.tex                        2 December 2010
%  available symbols are:
%  $\ast, \dag, \ddag, \S, \P, $\|$, $\ast\ast$, \dag\dag, \ddag\ddag ,\#
%
\collaboration{The D0 Collaboration\footnote{with visitors from
%{alton}
$^{a}$Augustana College, Sioux Falls, SD, USA,
%{burdin}
$^{b}$The University of Liverpool, Liverpool, UK,
%{haas,partridge}
$^{c}$SLAC, Menlo Park, CA, USA,
%{juste}
$^{d}$ICREA/IFAE, Barcelona, Spain,
%{luna-garcia}
$^{e}$Centro de Investigacion en Computacion - IPN, Mexico City, Mexico,
%{podesta-lerma}
$^{f}$ECFM, Universidad Autonoma de Sinaloa, Culiac\'an, Mexico,
and
%{weber}
$^{g}$Universit{\"a}t Bern, Bern, Switzerland.%
%{garcia-guerra}
%$^{?}$UPIITA-IPN, Mexico City, Mexico,
%{hooper}
%$^{?}$%Visitor from Bradley University, Peoria, IL, USA.
%{kozminski
%$^{?}$}%Visitor from Lewis University, Romeoville, IL, USA.
%{deceased}
%$^{\ddag}$%Deceased.
}} \noaffiliation
\vskip 0.25cm

\date{January 3, 2011}

\begin{abstract}
We present the first measurement of the color representation of the
hadronically decaying $W$ boson in $t\bar{t}$ events, from 5.3~\ifb\ of
integrated luminosity collected with the D0 experiment. A novel
calorimeter-based vectorial variable, ``jet pull,'' is used, sensitive to the
color-flow structure of the final state. We find that the fraction of
uncolored $W$ bosons is $0.56\pm 0.42\textrm{(stat+syst)}$, in agreement with
the standard model.
\end{abstract}

%Here are the relevant PACS numbers that we can quote in PRL
%(http://www.aip.org/pacs/pacs06/pacs0610.html):
\pacs{12.38.Qk, 12.38.Aw, 14.65.Ha}

\maketitle

Color charge is conserved in quantum chromodynamics (QCD), the theory that
describes strong interactions~\cite{qcd}. At leading order in the strong
coupling constant $\alpha_s$, color can be traced from initial partons to
final-state partons in high-energy hadron collisions. Two final-state partons
on the same color-flow line are ``color-connected'' and attracted by the
strong force. As these colored states hadronize, the potential energy of the
strong force between them is released in the form of hadrons. Thus, knowledge
of the color-connections between jets can serve as a powerful tool for
separating processes that otherwise appear similar. For example, in the decay
of a Higgs ($H$) boson to a pair of bottom ($b$) quarks, the two $b$ quarks
are color connected to each other, since the $H$ is uncolored (color
singlet), whereas in $g\rightarrow b\bar{b}$ background events, they are
color-connected to beam remnants because the gluon carries a color and an
anti-color (color-octet). We follow a recent
suggestion~\cite{Gallicchio:2010sw} for reconstructing these color
connections experimentally, using observables that can be modeled reliably by
available leading-log parton-shower simulations. The technique involves
measuring a vectorial quantity called ``jet pull,'' which represents the
eccentricity of the jet in the $\eta$-$\phi$ plane~\cite{d0coord} and the
direction of the major axis of the ellipse formed from the jet energy
pattern. Jets tend to have their pull pointing towards their color-connected
partner. For instance, in $H\rightarrow b\bar{b}$ events, the pulls of the
two $b$-jets tend to point towards each other, whereas in $g\rightarrow
b\bar{b}$ events, they point in opposite directions along the collision axis.

Verification of color flow simulation and jet pull reconstruction for both
color-singlet and color-octet configurations is interesting in its own
right~\cite{Sung:2009iq} and is needed before jet pull can be used in, e.g.,
$H\rightarrow b\bar{b}$ searches. Color-octet patterns can be studied in many
processes, such as $W$/$Z$ boson production in association with jets. A pure
sample of color-singlet hadronic decays is difficult to obtain at a hadron
collider, but $t\bar{t}$ events with an $\ell$+jets final state are good
candidates since they have a characteristic signature and contain two jets
from the decay of a $W$ boson, which is a color singlet. Each of the two
$b$-jets coming from the top quark decays is color-connected to one of the
beam remnants in a color-octet pattern.

In this Letter, we use data collected with the D0 detector~\cite{d0det} at
the Fermilab Tevatron $p\bar{p}$ collider, corresponding to 5.3~fb$^{-1}$ of
integrated luminosity, to present the first experimental results on the study
of jet pull, using $t\bar{t}$ events decaying to $\ell$+jets
($t\bar{t}\rightarrow WbW\bar{b} \rightarrow \ell \nu b j\bar{j} \bar{b}$,
where $\ell = e$,$\mu$). The object identification, event selection, and
simulated Monte Carlo (MC) events are the same as those used in the
$t\bar{t}$ cross section analysis~\cite{xsecnote}, except that looser
$b$-tagging criteria~\cite{Abazov:2010ab} are used to increase the statistics
of double $b$-tagged events. We obtain a $\approx 90\%$ pure $t\bar{t}$
sample by requiring an isolated lepton with $p_T>20$~GeV, missing transverse
energy \MET $>20$~GeV ($>25$~GeV for the $\mu$+jets channel), and at least
four jets, reconstructed with a midpoint cone algorithm~\cite{RunIIcone} of
radius 0.5, with $p_T>20$~GeV. At least one jet must have $p_T>40$~GeV, and
at least two jets must be identified as $b$-jets.
Table~\ref{tab:yields_ljets} shows the event yields for these selection
criteria.

\begin{table}
\begin{center}
\caption{Yields of events passing selections with exactly $4$ or $\ge5$ jets.
At least two $b$-tagged jets are required in the analysis, but the numbers of
events with zero or one $b$-tagged jet are also given. The number of
$t\bar{t}$ events is calculated using the cross section determined with this
data sample, $\sigma_{t\bar{t}}=8.50$~pb. Uncertainties include statistical
and systematic contributions. The total uncertainties are smaller than the
sum of individual uncertainties due to negative correlations between
samples.} \label{tab:yields_ljets}
\begin{tabular}{l  l  r@{ $\pm$ }l  r@{ $\pm$ }l  r@{ $\pm$ }l }
\hline \hline
 channel   & sample & \multicolumn{2}{c}{0 $b$-tags} &
 \multicolumn{2}{c}{1 $b$-tag } & \multicolumn{2}{c}{$\ge$ 2 $b$-tags } \\
\hline
$\ell$+4\,jets & $W$+jets   & 576 & 75 & 229 & 32 & 49 &  8 \\
          & Multijet   &  115 &  16 & 46  &  7 &  7 &  2 \\
          & $Z$+jets   &  42 &  6 & 16  &  3 &  4 &  1 \\
          & Other      &  31 &  4 & 19  &  2 &  9 &  1 \\
          & $t\bar{t}$ &  160 &  22 & 417 & 38 & 519 & 51 \\
          & Total      & 923 & 62 & 727 & 24 & 589 & 48 \\
          & Observed   & \multicolumn{2}{c}{923} & \multicolumn{2}{c}{743} &  \multicolumn{2}{c}{572} \\
\hline
$\ell$+$\ge$5\,jets & $W$+jets   &  60 & 22 & 26 & 11 & 7 &  3 \\
          & Multijet   &  17 &  3 & 12  &  2 & 3 &  1 \\
          & $Z$+jets   &  4 &  1 & 2  &  1 &  1 &  1 \\
          & Other      &  3 &  1 & 3  &  1 &  2 &  1 \\
          & $t\bar{t}$ &  34 &  6 & 90 & 13 & 132 & 17 \\
          & Total      & 118 & 19 & 132 & 7 & 145 & 15 \\
          & Observed   & \multicolumn{2}{c}{112} & \multicolumn{2}{c}{127} &  \multicolumn{2}{c}{156} \\
\hline \hline
\end{tabular}
\end{center}
\end{table}

\begin{figure}\centering
\setlength{\unitlength}{1.0cm}
\begin{picture}(6.0,4.5)
\includegraphics[width=6cm]{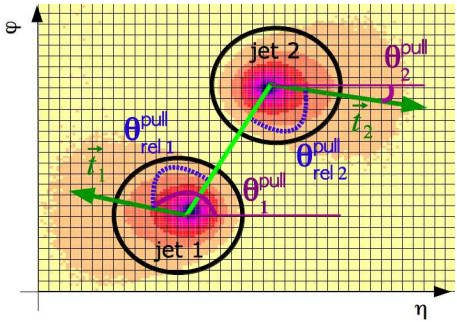}
\end{picture}
\caption{(Color online) Diagram showing two jets in the $\eta$-$\phi$ plane,
and the reconstruction of the jet pull vectors ($\vec{t}$), jet pull angles
($\theta^{\rm pull}$), and relative jet pull angles ($\theta^{\rm pull}_{\rm
rel}$).} \label{fig:pullrel}
\end{figure}

To extract the fraction of color-singlet hadronic $W$ boson decays, the data
are compared to both standard model $t\bar{t}$ MC (with a color-singlet $W$
boson) and an alternative model of $t\bar{t}$ with a hypothetical color-octet
``$W$'' boson decaying hadronically with identical properties except for its
color representation. The latter is simulated using the
\madgraph~(MG)~\cite{mad} event generator interfaced to \pythia~\cite{pythia}
for showering and hadronization. Simulated events are processed with a
\GEANT3-based~\cite{geant} detector simulation, overlaid with random data to
account for backgrounds, and reconstructed as data.

D0 uses three liquid-argon/uranium calorimeters to measure the energies of
particles: a central section (CC) covering $|\eta|$ up to $\approx 1.1$ and
two end calorimeters (EC) that extend coverage to $|\eta|\approx
4.2$~\cite{d0coord}, housed in separate cryostats~\cite{run1det}. In
addition, scintillators between the CC and EC cryostats provide sampling of
developing showers for $1.1<|\eta|<1.4$. There are approximately ten layers
in the radial direction (depending on $\eta$), generally composed of cells
spanning $0.1\times0.1$ in $\eta\times\phi$. The energy resolution is about
$15\%/\sqrt{E} \oplus 0.3\%$ (in GeV) for electrons and $50\%/\sqrt{E} \oplus
5\%$ for hadrons. Pileup energy from overlapping \ppbar\ interactions result
in about 0.5\% of cells having energy above the noise-limited energy
threshold ($\approx 50-500$ MeV, depending on layer and $\eta$). This energy
is roughly exponentially distributed, with a mean of $\approx 350$ MeV.

The pull is determined for each jet of a pair of reconstructed jets, using
the measured energies of the calorimeter cells (see Fig.~\ref{fig:pullrel}).
Each cell within $\Delta R=\sqrt{(\Delta\phi)^2+(\Delta\eta)^2}<0.7$ of the
$E_T$-weighted center of one of the jets of the pair ($\eta_d^{\rm jet}$,
$\phi^{\rm jet}$) is assigned to the jet nearer in $\Delta R$. The
contribution of each selected cell to the jet pull is $\vec{t}_{\rm cell} =
E_T^{\rm cell} |\vec{r}_{\rm cell}| \vec{r}_{\rm cell}$, where $\vec{r}_{\rm
cell}$ = ($\eta_d^{\rm cell}-\eta_d^{\rm jet}$, $\phi^{\rm cell}-\phi^{\rm
jet}$), and $E_T^{\rm cell}$ is the cell's transverse energy with respect to
the nominal center of the detector. The jet pull is $\vec{t}=\sum_{{\rm
cells},i}\vec{t}_i/E_T^{\rm jet}$. The polar angle of the jet pull,
$\theta^{\rm pull}$, is defined to be zero when pointing in the positive
$\eta$ direction along the beamline. A small correction to the jet pull is
made to account for the energy response and noise in the calorimeters as a
function of $\eta_d$, particularly in regions between the central and forward
cryostats. The angle of the jet pull direction relative to the line defined
by the centers of the jet pair ($\theta^{\rm pull}_{\rm rel}$) is also of
interest, as we expect color-connected jets to have pulls pointing towards
each other. The $\theta^{\rm pull}_{\rm rel}$ quantity is calculated for each
jet in the pair of highest-$p_T$ $b$-tagged jets ($b$ pair) and the pair with
highest $p_T$ which are not amongst the two highest $p_T$ $b$-tagged jets
($w$ pair).

To select events with a higher purity of properly identified jet pairs from
hadronic $W$ boson decays, we split the sample into events where the
invariant mass of the $w$-pair jets is consistent with the $W$ boson mass,
$|m_{jj}-M_W|<30$~GeV, and events where it is not. For the former, these two
jets are found to match the partons from the $W$ boson decay within $\Delta
R<0.5$ in 66\% of $t\bar{t}$ MC events with four jets and 46\% of events with
5 or more jets. In the latter case, additional gluon radiation in the initial
or final state leads to possible additional color configurations, diluting
the measurement.

\begin{figure}\centering
\setlength{\unitlength}{1.0cm}
\begin{picture}(5.0,3.0)
\put(-2.1,0.1){\includegraphics[width=4.7cm]{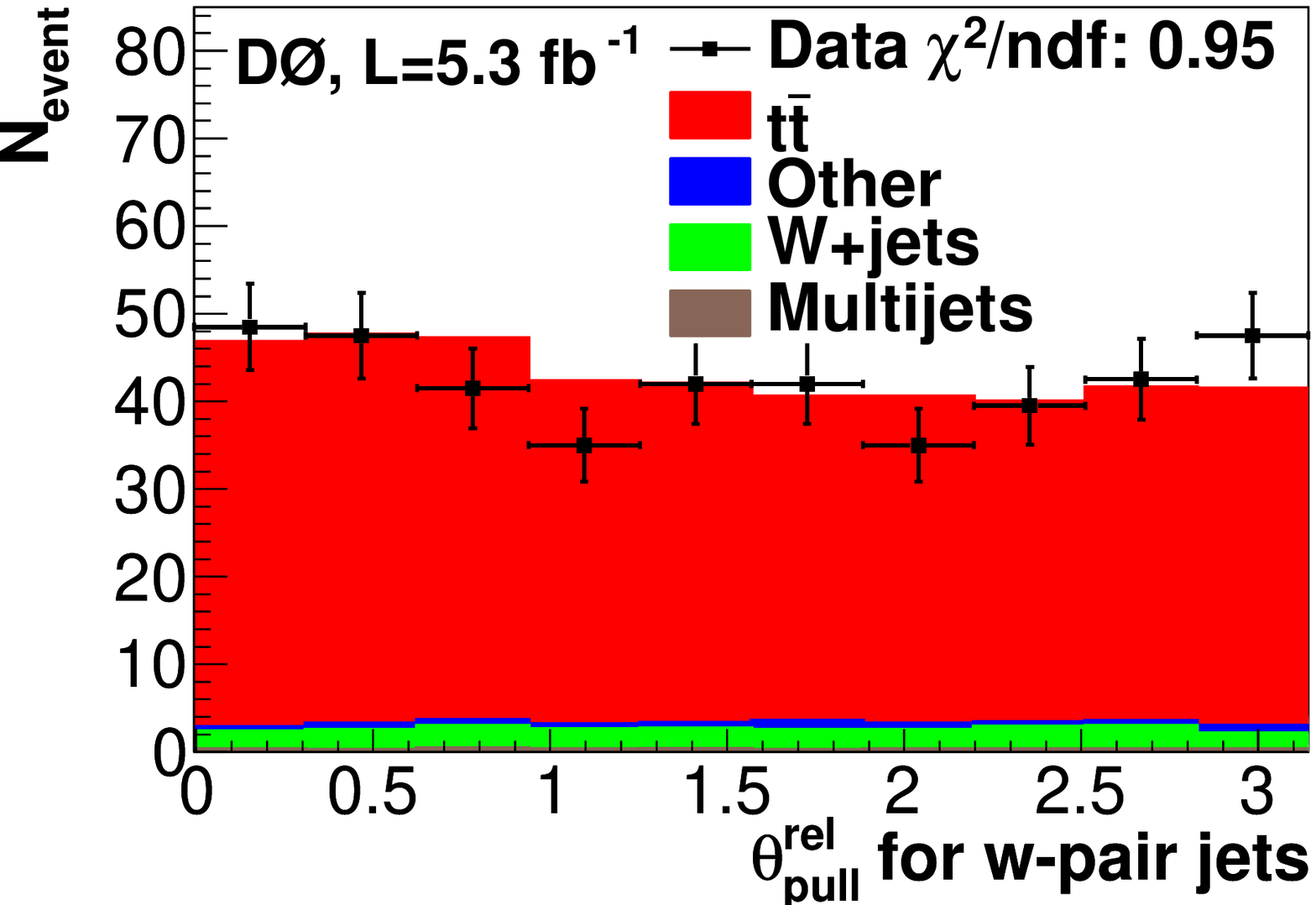}}
\put(2.45,0.1){\includegraphics[width=4.7cm]{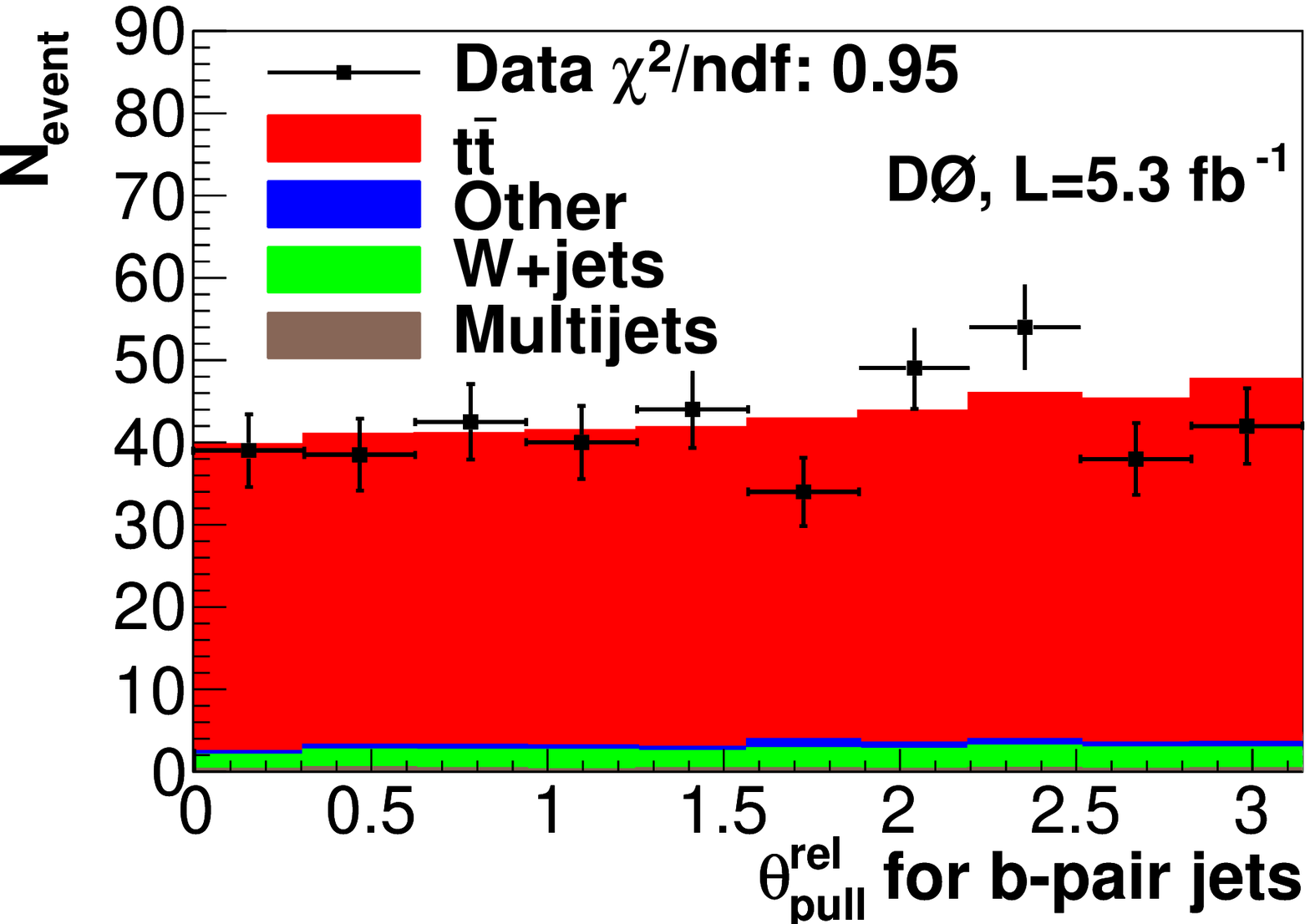}}
\put(-1.2,1.0){\rm \bf(a)} \put(3.45,1.0){\rm \bf(b)}
\end{picture}
\caption{(Color online) The average of the two jet $\theta^{\rm pull}_{\rm
rel}$ distributions for jets in pairing (a) $w$ and (b) $b$, in events with
exactly four jets, at least two $b$-tags, and the $M_W$ requirement on the
$w$-pair jets. The $\chi^2$/ndf compares the data to the total MC
distribution.} \label{fig:pair1_mw}
\end{figure}

\begin{figure}\centering
\setlength{\unitlength}{1.0cm}
\begin{picture}(5.0,3.0)
\put(-2.1,0.1){\includegraphics[width=4.7cm]{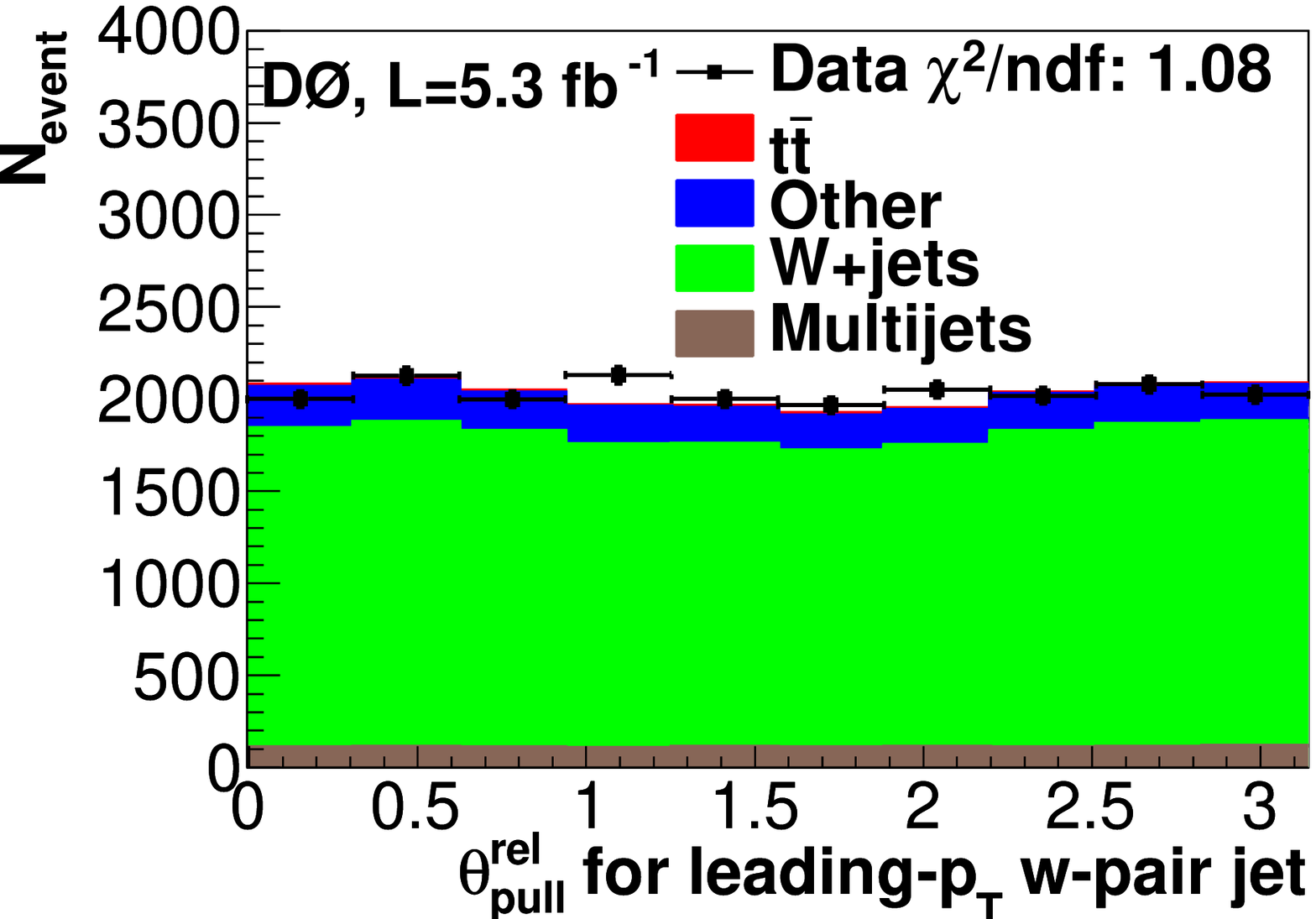}}
\put(2.45,0.1){\includegraphics[width=4.7cm]{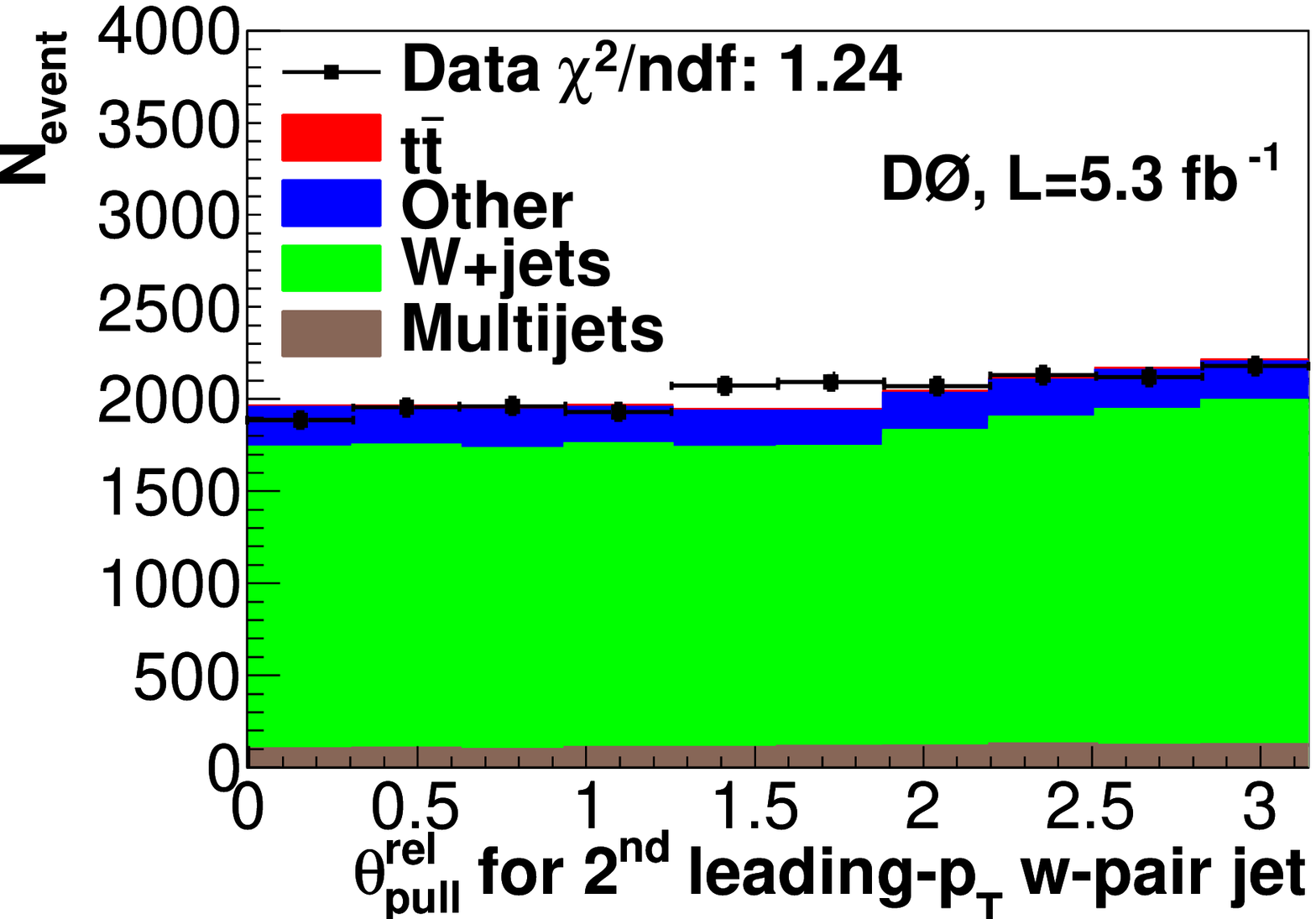}}
\put(-1.15,1.0){\rm \bf(a)} \put(3.5,1.0){\rm \bf(b)}
\end{picture}
\caption{(Color online) (a) Leading-$p_T$  and (b) second-leading-$p_T$ jet
$\theta^{\rm pull}_{\rm rel}$ distributions for $w$-pair jets, in events with
two jets and no $b$-tagged jets. The $\chi^2$/ndf compares the data to the
total MC distribution.} \label{fig:w2jets_0tag}
\end{figure}

Since the $w$-pair jets in $t\bar{t}$ events are often from the $W$ boson
decay, we expect them to be color-connected, thus the jet pulls should
generally point towards each other. We expect $b$-pair jets to have one of
the $b$-jets color-connected to the proton beam and the other to the
anti-proton beam, thus the jet pulls should be generally pointing away from
each other. This tendency is seen in data as shown in
Fig.~\ref{fig:pair1_mw}, with smaller $\theta^{\rm pull}_{\rm rel}$ in the
$w$ pair than in the $b$ pair. However, the jets in $w$ and $b$ pairs have
different kinematics, separation in the detector, and flavor. A direct
interpretation of the effects from color-flow is therefore not possible from
this comparison. Furthermore, there are detector and reconstruction effects
on jet pulls from overlapping jet pull cones, calorimeter noise and pileup,
and calorimeter response inhomogeneity. For instance, there would be fewer
cone overlaps if the jet pull was defined using only calorimeter cells within
$\Delta R<0.5$, producing on average smaller values for $\theta^{\rm
pull}_{\rm rel}$.  With this alternative definition the shape in
Fig.~\ref{fig:pair1_mw}(a) would peak more towards zero and that in
Fig.~\ref{fig:pair1_mw}(b) would be flatter. These effects are found to be
well-modeled by the simulation, and the jet pull definition based on the
$\Delta R<0.7$ cone gives a slightly improved singlet-octet separation. The
relative jet pulls $\theta^{\rm pull}_{\rm rel}$ in data are also found to be
well-modeled by simulation for other jet pairings, such as a random $w$-pair
jet and a random $b$-pair jet. In control samples consisting of events with a
leptonic $W$ boson decay, and two, three, or four jets, none identified as
$b$-jets, various jet pairings also have jet pulls that agree with
simulations. Figure~\ref{fig:w2jets_0tag} shows the $\theta^{\rm pull}_{\rm
rel}$ distributions for jets in a control sample with a leptonic $W$ boson
decay and two not-$b$-tagged jets.

To quantify the method's sensitivity to the color-flow structure
(color-singlet versus color-octet) for the hadronic $W$ boson decay, we fit
the data to two hypotheses: (i) standard model $t\bar{t}$ with a
color-singlet hadronically decaying $W$ boson (singlet MC) and (ii)
$t\bar{t}$ with a hypothetical color-octet ``$W$'' boson (octet MC). We
determine the fraction of events coming from color-singlet $W$ boson decay
($f_{\rm Singlet}$) using the fitting procedure from the D0 combined
$t\bar{t}$ cross section analysis~\cite{xsecnote}. We simultaneously measure
the $t\bar{t}$ cross section to avoid any possible influence of the
$t\bar{t}$ signal normalization on the $f_{\rm Singlet}$ measurement. The
discriminating variable used for the fit is derived from the $\theta^{\rm
pull}_{\rm rel}$ angles of the $w$-pair jets and depends on the $\Delta R$
between the two jets and their $\eta_d$. For events failing the $W$ mass
requirement, we do not split the regions further; for other events we split
the data sample according to the $\eta_d$ of the jets and $\Delta R$ between
the jets. For events where the two jets are highly separated ($\Delta R>2$),
we use the $\theta^{\rm pull}_{\rm rel}$ of the leading-$p_T$ jet. Little
discrimination is possible for these events, since the additional color
radiation is distributed over a large area of the calorimeter. When the two
jets are close ($\Delta R<2$) and $|\eta_d|<1.0$ for both jets, we use the
minimum $\theta^{\rm pull}_{\rm rel}$ of the two jets. This is the most
sensitive region, and the jet pull is accurately reconstructed in the central
calorimeter due to less pileup energy and uniformity of response. Otherwise,
if $|\eta_d|$ of the leading-$p_T$ jet is $<1.0$ ($>1.0$), the $\theta^{\rm
pull}_{\rm rel}$ of the leading-$p_T$ (second-leading $p_T$) jet is used.

Table~\ref{tab:systs_fsinglet_standard2} lists the contribution of each
non-negligible source of systematic uncertainty on $f_{\rm Singlet}$. For all
but the theoretical cross sections, MC statistics, and normalization of the
$W$+heavy flavor jets background uncertainties, we apply the systematic
uncertainties just to the \ttbar\ signal sample and ignore the effect on
background, as the purity of the \ttbar\ sample is high. To estimate the
possible systematic shift of the $\theta^{\rm pull}_{\rm rel}$ distribution
due to the different energy scale and noise of the calorimeter cells between
data and MC as a function of $\eta_d$, we apply $\pm$50\% of the jet pull
$\eta$ correction and take the resulting difference in shape as the
systematic uncertainty for jet pull reconstruction. This covers the
differences in the average $\theta^{\rm pull}$ when comparing data and MC
control samples. We also study systematic uncertainties as
in~\cite{xsecnote}, the main ones being from the jet energy scale, jet energy
resolution, $b$-tagging efficiency, and lepton misidentification. Additional
systematic uncertainties on $\theta^{\rm pull}_{\rm rel}$ are assessed to
account for possible differences between MC and data related to the modeling
of underlying event, hadronization, and jet showering. To estimate the
variation due to these possible mis-modelings, we compare $\theta^{\rm
pull}_{\rm rel}$ distributions in events simulated with \pythia\, to those
with \alpgen~\cite{alpgen} or \mcatnlo~\cite{mcnlo}, and showering with
\herwig~\cite{herwig}. We also do the comparisons for various \pythia\,
parameters for underlying event and
color-reconnection~\cite{Sjostrand:2004pf}, such as tunes APro and
NOCR~\cite{tunes}. When deriving $f_{\rm Singlet}$ from the fit, we use the
maximal variation obtained with the different $\theta^{\rm pull}_{\rm rel}$
distributions as an estimate of the systematic uncertainty.

\begin{table}
\begin{center}
\caption{The one standard deviation ($\sigma$) variation of $f_{\rm Singlet}$
from main systematic uncertainties. The total systematic uncertainty includes
all uncertainties, summed in quadrature.}
\label{tab:systs_fsinglet_standard2}
\begin{tabular}{ccc} \hline \hline
 Source  &     $+1\sigma$    &    $-1\sigma$   \\ \hline
               Singlet/octet MC shapes  & 0.188 & $-0.188$ \\
                 Jet pull reconstruction & 0.100& $-0.093$  \\
                  Jet energy resolution & 0.033  & $-0.013$  \\
                  Vertex confirmation & 0.028  & $-0.029$  \\
                 \pythia\, tunes &  0.023 & $-0.025$ \\
                  Jet energy scale & 0.024  &  $-0.009$ \\
                  Jet reconstruction and identification & 0.017  & $-0.017$ \\
                    $t\bar{t}$ modeling &  0.014  & $-0.033$  \\
                  Event statistics for matrix method & 0.009 & $-0.010$ \\
                  Other Monte Carlo statistics & 0.009 & $-0.007$ \\
                  Multijet background & 0.006 & $-0.007$ \\
      \hline
                   Total systematic & 0.222 & $-0.218$ \\
\hline \hline
\end{tabular}
\end{center}
\end{table}

Since the results are statistically limited and the analysis does not as yet
provide sufficient sensitivity for a definitive observation of color-flow, we
set limits on $f_{\rm Singlet}$ using the likelihood ratio ordering scheme of
Feldman and Cousins~\cite{feldmancousins}. We follow the same approach used
for the simultaneous extraction of the ratio of branching fractions and the
$t\bar{t}$ cross section~\cite{rb} and generate ensembles of
pseudo-experiments for different values of $f_{\rm Singlet}$ between 0 and 1,
with the $t\bar{t}$ cross section fixed to the measured value. We then vary
the systematic uncertainties using Gaussian distributions and perform the fit
as for the measurement on data. Statistical uncertainties are incorporated by
smearing the measured value for each pseudo-experiment with the uncertainty
determined in data. We use the nuisance parameters method where the
expectation is fit to the data, for a variation of the initial prediction
within the systematic uncertainties, allowing also the central result to
change~\cite{xsecnote}. Other methods give compatible results.

\begin{figure}\centering
\includegraphics[width=8cm]{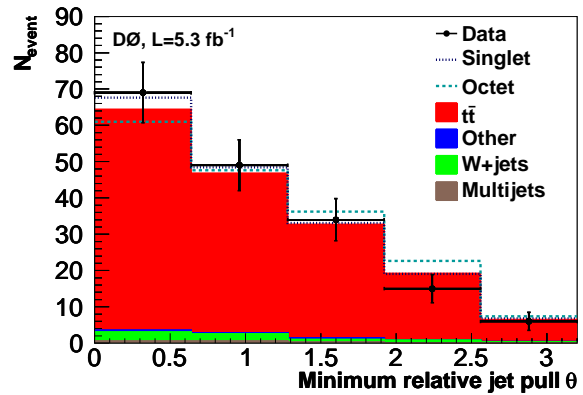}
\caption{(Color online) The discriminating color-flow variable, the minimum
$\theta^{\rm pull}_{\rm rel}$ for the $w$-pair jets, for events passing the
$M_W$ requirement, with $\Delta R<2$, and $\eta_d<1.0$ for both jets. The
$t\bar{t}$ MC shape is obtained using the measured value of $f_{\rm
Singlet}$.} \label{fig:measured_jetpullrel_lead}
\end{figure}

\begin{figure}\centering
\includegraphics[width=8cm]{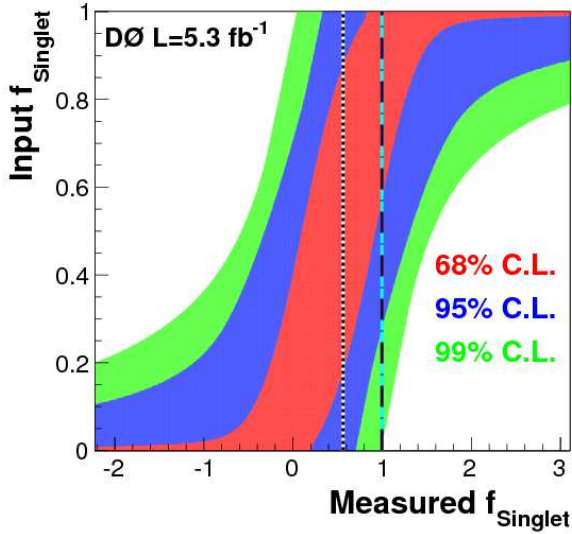}
\caption{(Color online) Expected C.L. bands for $f_{\rm Singlet}$. The
measured value is shown on the horizontal axis, and the input value on the
vertical axis. The wide-dashed line shows the expected value and the
black-white fine-dashed line indicates the measured value of $f_{\rm
Singlet}$.} \label{fig:measured_limit}
\end{figure}

We measure $f_{\rm Singlet} = 0.56 \pm 0.42\ [\pm 0.36\textrm{(stat)} \pm
0.22\textrm{(syst)}]$ and $\sigma_{t\bar{t}} = 8.50^{+0.87}_{-0.76}$ pb,
consistent with our dedicated cross section measurement~\cite{xsecnote}.
Figure~\ref{fig:measured_jetpullrel_lead} shows the distribution for one of
the regions of the discriminating color-flow variable, using the measured
$t\bar{t}$ cross section and measured $f_{\rm Singlet}$. The expected 99\%
C.L. and 95\% C.L. limits are $f_{\rm Singlet}>0.011$ and $f_{\rm
Singlet}>0.277$ respectively, corresponding to an expected sensitivity to
exclude $f_{\rm Singlet}=0$ of about three standard deviations, based on
pseudo-experiments. The 68\% C.L. allowed region from data is $0.179 < f_{\rm
Singlet} <0.879$. Figure~\ref{fig:measured_limit} shows the expected 68\%,
95\%, and 99\% C.L. bands for $f_{\rm Singlet}$.

In summary, we have presented the first study of color flow in $t\bar{t}$
events, with the method of jet pull, using 5.3~\ifb\ of D0 integrated
luminosity. The standard model MC predictions are found to be in good
agreement with data, for both the jets from the hadronically decaying $W$
boson, which should be in a color-singlet configuration, and the $b$-tagged
jets from the top quark decays, which should be in a color-octet
configuration. To quantify our ability to separate singlet from octet
color-flow, we measured the color representation of the hadronically decaying
$W$ boson and found $f_{\rm Singlet}=0.56\pm 0.42\textrm{(stat+syst)}$, while
the expected 95\% C.L.\ limit was $f_{\rm Singlet}>0.277$. The ability to use
color flow information experimentally will benefit a wide range of
measurements and searches for new physics.

We thank Jason Gallicchio, Matthew Schwartz, Steve Mrenna, Peter Skands, and
Jay Wacker for discussions and guidance.
% acknowledgement.tex                             2 December 2010
%
We thank the staffs at Fermilab and collaborating institutions,
and acknowledge support from the
DOE and NSF (USA);
CEA and CNRS/IN2P3 (France);
FASI, Rosatom and RFBR (Russia);
CNPq, FAPERJ, FAPESP and FUNDUNESP (Brazil);
DAE and DST (India);
Colciencias (Colombia);
CONACyT (Mexico);
KRF and KOSEF (Korea);
CONICET and UBACyT (Argentina);
FOM (The Netherlands);
STFC and the Royal Society (United Kingdom);
MSMT and GACR (Czech Republic);
CRC Program and NSERC (Canada);
BMBF and DFG (Germany);
SFI (Ireland);
The Swedish Research Council (Sweden);
and
CAS and CNSF (China).


\begin{thebibliography}{99}
\bibitem{d0coord}D0 uses a right-handed coordinate system, with the $z$-axis
pointing in the direction of the proton beam and the $y$-axis pointing
upwards. The azimuthal angle $\phi$ is defined in the $xy$ plane and is
measured from the $x$-axis. The pseudorapidity is defined as $\eta =
-\ln[\tan(\theta/2)]$, where $\theta$
% = \arctan(\sqrt{x^2+y^2}/z)$.
is the polar angle. Detector $\eta$ ($\eta_d$) is the $\eta$ of an object
measured from the nominal detector center.
\end{thebibliography}
\end{document}